\documentclass[preprint]{article}      

\usepackage{graphicx}
\usepackage{amssymb}

\title{Computer Administering of the Psychological Investigations: Set-relational Representation}

\author{Krasimir Yordzhev\thanks{South-West University, Blagoevgrad, Bulgaria, Faculty of Mathematics and Natural Sciences, Email: yordzhev@swu.bg} \and Ivelina Peneva\thanks{South-West University, Blagoevgrad, Bulgaria,Faculty of Philosophy, Department of Psychology,  Email: ivelina\_peneva@swu.bg} }

\date{}

\begin{document}

\maketitle

\begin{abstract}
Computer administering of a psychological investigation is the computer representation of the entire procedure of psychological assessments - test construction, test implementation, results evaluation, storage and maintenance of the developed database, its statistical processing, analysis and interpretation. A mathematical description of psychological assessment with the aid of personality tests is discussed in this article. The set theory and the relational algebra are used in this description. A relational model of data, needed to design a computer system for automation of certain psychological assessments is given. Some finite sets and relation on them, which are necessary for creating a personality psychological test, are described. The described model could be used to develop real software for computer administering of any psychological test and there is full automation of the whole process: test construction, test implementation, result evaluation, storage of the developed database, statistical implementation, analysis and interpretation. A software project for computer administering personality psychological tests is suggested.

\end{abstract}

\textbf{Keywords}: {Computer administering, Computer testing, Computer Psychodiagnostic, Mathematical modeling, Set theory, Relational algebra, Databases  }

\textbf{MSC[2010] code:} { 91E45, 68P15}

\section{General concepts and definitions}
\label{Introduction}

\emph{Psychological assessment} will be called any psychological testing made with the help of a preliminary prepared test - a list of questions or statements, that the assessed person or group of people has to answer or to give their opinion for. The separate parts of the test are called \emph{items}.

We will call \emph{computerized psychological assessment} any psychological testing, where one or several phases of the testing are made with the use of a computer.

\emph{Computer administering of a psychological investigation} is the computer representation of the entire procedure of psychological assessments - test construction, test implementation, results evaluation, storage and maintenance of the developed database, its statistical processing, analysis and interpretation.

\emph{Personality questionnaires} are those psychological tests, which are purposed for description and evaluation of the characteristics of conative (behavioral), emotional and motivation sphere, the interpersonal relations and attitudes of an individual \cite{Sidorenko}. It's typical for the Personality questionnaires (in contrast to Achievement tests or Intelligence tests) that the items are questions or statements, for which answers the respondent has to report certain information concerning himself, his experience and relations. The format of the answers to items is also specific - most often they are described with the help of à finite set of preliminary known answers or statements, which we will mark by $Ans$. In our examinations we will accentuate mainly on finite sets of possible responds to each item of the test. For example  $Ans=$\{''\textsc{Yes}'', ''\textsc{No}''\},  $Ans=\{$''\textsc{True}'', ''\textsc{False}''\},  $Ans=$\{''\textsc{I like it}'', ''\textsc{I don't like it}''\},  $Ans=$\{''\textsc{Often}'', ''\textsc{Sometimes}'', ''\textsc{Never}''\},  $Ans=$\{''\textsc{True}'', ''\textsc{I don't know}'', ''\textsc{False}''\}, $Ans =$\{''\textsc{Agree}'', ''\textsc{I'm not sure}'', ''\textsc{Disagree}''\}. Items with rating scales are also used in practice, but we will examine just rating scales that represent definite and consequently discrete set of real numbers.

For more details about basic terms in personality psychological testing see for example in \cite{Anastasi,Miller}.

There are different ways to structure, process and store data in a software product. Data and data links are abstraction of facts and relations from the real world \cite{Hernandez}. Very often this abstraction is rather complex and requires a special mathematical model for its description - a \emph{data model}. The use of one or other model means that in certain information system are chosen different principles for data structuring or data operation. The most wide-spread data model nowadays is the \emph{relational model}, suggested for the first time by E. F.  Codd at the beginning of the 1970s of the last century and described in a set of articles, one of the earliest is \cite{Codd}. For this article E. F. Codd won a prestigious A. Turing award of the American Association for Computing Machinery in 1980.

Let the family of sets $D=\{ D_1 ,D_2 ,\ldots ,D_m \}$, which we will call \emph{domains}, be given. Let's examine the Cartesian product:
$$
\begin{array}{c}
W=D_{i_1} \times d_{i_2} \times \cdots \times D_{i_n} = \\
=\left\{ \langle x_1 ,x_2 ,x_n \rangle  \; |\; x_{i_k}\in D_{i_k} ,\; D_{i_k} \in D,\; k=1,2,\ldots ,n\right\}
\end{array}
$$

It is possible for some $s$  and  $t$,  $D_{i_s} = D_{i_t}$. Each subset
$$\rho\subseteq W$$
is called  $n$-ary \emph{relation on} $D$. From a practical point of view the finite relations are of interest, i.e. all possible finite subsets of $W$  and therefore in the present piece of writing "relation" will mean "finite relation" (besides the opposite is explicitly emphasized), no matter it is supposed that the sets $D_i$, $i=1,2,\ldots m$  could possibly be infinite (for example infinite sets of real numbers).

Let's mark with $R$  the set of all  $n$-ary relations formed by $n$, $n=1,2,\ldots$  (not definitely different) sets (domains) from the family $D$.

Let $\rho\in R$  and let $r^i =\langle  r_1^i ,r_2^i ,\ldots , r_n^i \rangle $  be the  $i$-th element of $\rho$.  $r^i$ is called the  $i$-th \emph{record} of the relation $\rho$. In this case  $i$ is an index, not an exponent. The component $r_j^i$  of $r^i$  is called \emph{value of the $j$-th attribute in the  $i$-th record of $\rho$}. The values of the  $j$-th attributes of all records of $\rho$ make the $j$-th \emph{field} of $\rho$. From the definition of the term field follows that the elements of  $j$-th field can receive values from one sole domain .

In the set $R$ of all relations in $D$ in certain circumstances it is possible to define various operations - union, intersection, subtraction, complement, projection, composition, indexing, sorting, etc. Relationships responding to certain conditions are possible between the separate attributes. Thus  $R$ together with the introduced operations and relationships between attributes turns into algebra, called \emph{relational algebra}. Relational algebra is in the base of relational data model. The database management systems, which have the relational model in their bases, are also called \emph{relational databases}. Basic knowledge in the field of the theory of relational algebras can be obtained in \cite{Maier}.

Each relation $\rho\in R$  could be visually presented like a rectangular table in which the  $i$-th row is the  $i$-th record, and the $j$-th column is the $j$-th field of $\rho$. This correspondence is one-to-one, i.e. the so built table completely determines the relation presented by it. Due to that reason and to help each user easily understand the main notions of relational algebra, in the most of software manuals for relational database they talk about and operate mostly with the notion \emph{table} as a synonym of the notion relation in the family of domains \cite{Hernandez}, never mind that from a practical point of view there could be a greater number of table types and not every table could present concrete relation, i.e. not every type of table could be used in a software for relational database. For example the calendar is a rectangular table, which cannot be identified with the above defined term relation.

\section{Sets of values and relation on them, which are necessary for creating a personality psychological test} \label{section2}

To create a Personality psychological test, using the help of a computer, it is necessary to define and specify the following sets:

2.1. The set $Qst$  is compound of questions or statements, presented for answer or opinion to the tested individual or group of people from the research psychologist. The elements of $Qst$ are the items of the test.

The presenting order of the items has a significant meaning for the psychologist, as the sequence of the discussed items has influence to answers and therefore is important for drawing the final conclusion and test results interpretation. In this relation arises the need of next set:

 2.2. Set  $Z_m =\{ 1,2,\ldots ,m\}$, where $m=|Qst |$  is the cardinal number of the elements of set  $Qst$. Each element of the set  $Z_m$ is a unique number of item in the set   $Qst$.

2.3. There is a bijective mapping  $\varphi \subset Qst \times Z_m$ between the sets $Qst$  and  $Z_m$, and the author of the psychological test for each $q\in Qst$  has to define very carefully the image $k=\varphi (q)$  of the element $q$  in the mapping $\varphi$. In this case $k$  is a number of the item $q$  and determines the order of items presentation to the tested individual. As it was above emphasized, determining the number of each item is important for the final conclusion and all that is in the competence of the psychologist - author of the test. In addition the number must be in the interval $[1,m]$  of natural numbers. In this sense in a database management system, needed for computer Personality psychological test (Tests generator) development when deleting an item or inserting a new one between two existing items, an automatic items renumbering has to be provided. In this sense the field  $Z_m$ is quite different from the auto increment field envisaged in a number of database management systems, which serves a primary key. For fields of that type the above mentioned operations division and item insertion are not followed by renumbering, even more it's not allowed to make changes in the primary key value.

2.4. A finite set $Ans$  of possible answers to the items (see section \ref{Introduction}).

2.5. A set  $Ctg$ of \emph{psychological categories} (\emph{Personality characteristics}), which are subject of analysis and evaluation concerning the assessed individual or group of people with the aid of the items from the set $Qst$   and the concrete answer that is chosen. It is not obligatory for each item to be related to given psychological category.

2.6. We are examining the finite family of subsets of $Qst$
$$
T=\left\{ T_c \subseteq Qst\; |\; c\in Ctg\right\}
$$
with the element $q\in Qst$  which belongs to the subset   $T_c$, if and only if the item $q$  has a relation to the category $c\in Ctg$  in the correspondent psychological assessment. Obviously $\displaystyle \bigcup_{c\in Ctg} T_c \subseteq Qst$.  If $q\in Qst$  and $\displaystyle q\not\in \bigcup_{c\in Ctg} T_c$, then the item $q$  will not effect the entire test and will drop out.

2.7. Let  $Ans =\{ a_1 ,a_2 ,\ldots ,a_k \}$. For every $a_i \in Ans$, $i=1,2,\ldots ,k$  we put together the set of numbers  $S_i$. Each element  $s\in S_i$, where $a_i \in Ans$, $i=1,2,\ldots ,k$  represents an assessment numerical value of the psychological categories from the set $Ctg$  on condition that the assessed person could possibly respond to a random item with an answer or a statement  $a_i \in Ans$. Quite often these assessment values are 1 or 0. For example, if $|Ans| =2$  and we assume that the assessment value for each positive answer is equal to 1 and for each negative answer is 0, then we can easily calculate the total number of positive answers.
We are building the  set:
$$
Scl =\prod_{a_i \in Ans} S_i =\left\{ \langle s_1 ,s_2 ,\ldots ,s_k \rangle \; |\; s_i \in S_i ,\; i=1,2,\ldots ,k,\; k=|Ans| \right\}
$$

2.8.
For each psychological category $c\in Ctg$  is defined the function
$$f_c \; :\; Qst \to Scl$$
with the aid of which the psychologist evaluates the assessed person regarding the category $c\in Ctg$. Such function is called a \emph{scale for the psychological category} $c\in Ctg$. It's a common practice the name of the scale to be exactly the same as the name of the psychological category. The set of functions
$$\left\{ f_c \; |\; c\in Ctg \right\} ,$$
composed by all scales for the examined psychological categories is called a \emph{scale of the personality psychological test}.

What is the assessment of the different items, regarding the correspondent answers, i.e. how the functions $f_c \; :\; Qst \to Scl$, $c\in Ctg$ are defined and what is the influence of this to the summary assessment of the test implementation can be estimated after wide psychological and statistical investigations \cite{Anastasi,Miller,Nasledov,Sidorenko}.

2.9. Relation (table)
$$
Bnd = \left\{ \langle c,q,f_c (q)\rangle \; |\; c\in Ctg , q\in Qst \right\} \subset Ctg \times Qst \times Scl ,
$$
where  $Ctg$, $Qst$  and $Scl$  are sets, which are defined correspondingly  in subsections 2.5, 2.1 and 2.7, and the relation  $f_c$, $c\in Ctg$  is a function defined in subsection 2.8 (scale of psychological category  $c$). Every record in this relation contains the number of the scores we can take when the person does the corresponding answer $a\in Ans$  according the scale $f_c (q)$  for category $c\in Ctg$  and item $q\in Qst$.

2.10. Let $c\in Ctg$  and let
$$
m_c = \min \sum_{q\in Qst} \mu \left( f_c \left( q\right)\right) ,
$$
where
$$
\mu (\langle  x_1 ,x_2 ,\ldots ,x_k \rangle ) =\min (x_1 ,x_2 ,\ldots , x_k ) .
$$

In other words $m_c$  is the minimum assessment value (minimum score), which could possibly be obtained in an arbitrary testing, related to the category $c\in Ctg$  and depending on the correspondent scale  $f_c \; :\; Qst \to Scl$.

Analogously we specify the maximum assessment value
$$
M_c = \max \sum_{q\in Qst} \nu \left( f_c \left( q\right)\right) ,
$$
where
$$
\nu (\langle  x_1 ,x_2 ,\ldots ,x_k \rangle ) =\max (x_1 ,x_2 ,\ldots , x_k ) .
$$

Obviously  $m_c <M_c$. We examine the finite sequence of numbers
$$
m_c =bc_0 <bc_1 <\cdots <bc_{l_c -1} <bc_{l_c} =M_c.
$$

As we denote  $l_c$ we emphasize the fact that the number of the members in the sequence depends on the category $c\in Ctg$. We are separating the interval $[m_c ,M_c ]$  to the subintervals  $[bc_0 ,bc_1 ]$,  $(bc_1 ,bc_2 ]$,  $(bc_2 ,bc_3 ]$,...,  $(bc_{l_c -1} ,bc_{l_c} ]$. Let's denote with $D_c$ the set of all intervals of that type $\Delta c_1 =[bc_0 ,bc_1 ]$ and  $\Delta c_i = (bc_{i-1} ,bc_i ]$,  $i=2,3,\ldots ,l_c$, corresponding to the category  $c\in Ctg$.

The number $l_c$  and each of the intervals $\Delta c_i$, $i=1,2,\ldots l_c$   should be a result of profound psychological investigations.

Let denote by
$$
D=\bigcup_{c\in Ctg} D_c
$$
the set of all intervals of numbers, which could be useful for certain personality psychological test.

2.11. For each of the intervals  $\Delta c_i$, $i=1,2,\ldots ,l_c$, $c\in Ctg$  defined in subsection 2.10, we create text  $tc_i$,  $i=1,2,\ldots ,l_c$, corresponding to the text given by a professional psychologist about the psychological condition of the assessed person regarding the psychological category  $c\in Ctg$, in case that the \emph{total summary} (the sum of  all concrete evaluation  for the corresponding item for the implementing test according to  the scale  $f_c$)  belongs to the interval  $\Delta c_i$. Thus for each  we obtain the sets of texts
$$
T_c =\left\{ tc_i \; |\; i=1,2,\ldots ,l_c \right\}
$$

2.12. Let  $c\in Ctg$. By the things we have mentioned above it follows that there exists functional dependency
$$
g_c \; :\; D_c \to T_c
$$

Let
$$
G_c =\left\{ \langle \Delta c_i ,tc_i \rangle \; |\; g_c (\Delta c_i )=tc_i ,\; i=1,2,\ldots , l_c \right\}
$$
is the corresponding binary relation of this dependency (graph of the function  $g_c$). The functional dependency $g_c : D_c \to T_c$  is called \emph{interpretation} of the test, concerning the psychological category $c\in Ctg$.

We denote by
$$
T=\bigcup_{c\in Ctg} T_c
$$
the set of all interpretations, which can be given of the psychological examiner after the implementation of a given personality psychological test.

2.13. At the end we define the relation
$$
Ntp =\left\{ \langle  c, \Delta c_i , g_c (\Delta c_i ) \rangle \; |\; c\in Ctg ,\; \Delta c_i \in D_c \right\} \subset Ctg \times D\times T
$$
where $D_c$  and $D$  are the sets of number intervals, we have already looked at 2.10., and $g_c$, $c\in Ctg$  and $T$  are correspondingly the functions and the set of interpretations, defined in 2.12.

The sense of each record in $Ntp$ is as follows: For each psychological category $c\in Ctg$ when the test is implemented if the assessed person obtain $r$ number of points (obtained assessment value for the category), then we check to which interval  $\Delta c_i \in D_c$, $c\in Ctg$ the number $r$ belongs and we give the expert interpretation of the test accordingly with the category $c\in Ctg$ and the function $g_c : D_c \to T_c$  defined in 2.12.

\section{A  software project for computer administering personality psychological tests}

Because of all that is written above, we consider that it is appropriate a computer system for administering personality psychological tests to contain three basic and several contributive, relatively independent modules. How do these separate modules work is shown on a diagram in Figure 1.
\begin{figure}[h]
\begin{center}
\begin{picture}(310,70)

\put(0,0){\framebox(70,30)}
\put(35,21){\makebox(0,0){Contributive}}
\put(35,9){\makebox(0,0){Modules}}

\put(0,40){\framebox(70,30)}
\put(35,55){\makebox(0,0){Generator}}

\put(110,20){\framebox(70,30)}
\put(145,35){\makebox(0,0){Executor}}

\put(200,20){\framebox(110,30)}
\put(255,41){\makebox(0,0){Modules for}}
\put(255,29){\makebox(0,0){Statistic Processing}}

\put(70,55){\line(1,0){20}}
\put(90,55){\line(0,-1){15}}
\put(90,40){\vector(1,0){20}}

\put(70,15){\line(1,0){20}}
\put(90,15){\line(0,1){15}}
\put(90,30){\vector(1,0){20}}

\put(180,35){\vector(1,0){20}}
\end{picture}
\end{center}
\caption{}\label{fig1}
\end{figure}

3.1. Module ''Generator''. This module will create files, containing sets and relations described in section \ref{section2} for a concrete personality psychological test.  At the moment we do not take an account of the data form, corresponding to the concrete database management system (Oracle Database, dBase, Paradox, FoxPro, MS Access, SQL, etc.).  Concrete realization of the module ''Generator'' is described in \cite{Yordzhev}.

3.2. One or several contributive modules. They serve for additional forming of a concrete computer test - creating starting window, enter a preliminary instruction, description of ''demographic field'' (if it is necessary) and others. The demographic field is a set of different data, characterizing the examined person as what is his/her sex, age, education, work, settlement, etc.

3.3. Module ''Executor''. This is a computer program, which reads the data from files and makes a concrete computer testing. From psychological point of view it is desirable the interface of this program to be as simple as possible. It is enough in the main window to appear the text of the serial item of the set  $Qst$. The order they appear is according to the bijective mapping between the elements of the set $Qst$  and the set  $Z_m =\{ 1,2,\ldots ,m\}$, $m=|Qst|$  in ascending order of the elements in  $Z_m$, described in subsection 2.3. By request just before the program finishes working appears a window ''Interpretation'', in which are described the conclusions obtained from analyzing the data of the concrete testing according to the relation $Ntp$ described in subsection 2.13. The data will be saved in a file, because we will need it for the coming modules of the system for computer administering of psychological tests.

3.4. Modules for statistic processing. They serve for automated statistic processing of the data, which is obtained when the module ''Executor'' is used many times, testing many persons. We can use universal programs for statistic process, which are often used in practice, such as SPSS, STATISTICA, MatLab, Maple, MS Excel and others. Certainly as we have in mind the specific features of the data, obtained in a result of a computer psychological testing, we consider that it is advisable to create specific software for the concrete statistic processing, which has to read and process the data, obtained when the module ''Executor'' works. This will lead to maximum automation of the psychological examinations

\bibliographystyle{unsrt}

\bibliography{SetRelRep}

\begin{thebibliography}{1}

\bibitem{Sidorenko}
E.~Sidorenko.
\newblock {\em Methods for mathematical processing in psychology}.
\newblock Rech, Sankt Petersburg, 2000.

\bibitem{Anastasi}
A.~Anastasi and S.~Urbina.
\newblock {\em Psychological testing}.
\newblock Prentice-Hall, 1997.

\bibitem{Miller}
S.~A. Miller.
\newblock {\em Developmental research methods}.
\newblock Prentice-Hall, Englewood Cliffs, NJ, 2 edition, 1998.

\bibitem{Hernandez}
M.~J. Hernandez.
\newblock {\em Database Design for Mere Mortals}.
\newblock Addison Wesley, 2 edition, 2003.

\bibitem{Codd}
E.~F. Codd.
\newblock A relational model of data for large shared data banks.
\newblock {\em Communication of the Association for Computing Machinery},
  13(6):377--387, 1970.

\bibitem{Maier}
D.~Maier.
\newblock {\em The theory of relational databases}.
\newblock 1983.

\bibitem{Nasledov}
A.~Nasledov.
\newblock {\em Mathematical methods for psychological assessments - analysis
  and data interpretation}.
\newblock Sankt Petersburg, 2004.

\bibitem{Yordzhev}
K.~Yordzhev, I.~Peneva, and B.~Kirilieva-Shivarova.
\newblock A relational model of personality psychological tests.
\newblock In {\em Mathematics and natural science, FMNS-2009}, volume~1, pages
  69--77. South-West University, Blagoevgrad, Bulgaria, 2009.

\end{thebibliography}

\end{document}